# Ultraluminous Starbursts in Major Mergers

J. Christopher Mihos and Lars Hernquist[1]

Board of Studies in Astronomy and Astrophysics,

University of California, Santa Cruz, CA 95064

*hos@lick.ucsc.edu, lars@lick.ucsc.edu*

## ABSTRACT

We use numerical simulation to investigate the triggering of starbursts in merging disk galaxies. The properties of the merger-driven starbursts are sensitive to the structure of the progenitor galaxies; specifically, to the amount of material in a dense central bulge. Galaxies without bulges develop bars shortly after their first close passage, driving significant gas inflow and subsequent starbursts in the centers of the galaxies. These starbursts significantly deplete the star-forming gas, so that only relatively weak starbursts arise during the final merger. By contrast, models of galaxies with central bulges show that a bulge acts to stabilize the galaxies against inflow and starbursts until the galaxies actually merge. At this time, strong dissipation leads to the formation of a massive central gas mass and an ensuing star formation rate two orders of magnitude greater than that in our isolated disk models. These starbursts are very short in duration, typically $\sim 50$ Myr, suggesting that the rarity of ultraluminous infrared galaxies is a result of their being in a very short evolutionary phase, rather than special and rare formation conditions. The fact that these mergers display many of the properties of ultraluminous infrared galaxies – tidal features, double nuclei, massive compact gas concentrations, and extreme levels of starburst activity – suggests that merger-driven starbursts can explain the emission from many ultraluminous infrared galaxies without an active nucleus.

*Subject headings:* galaxies:interactions, galaxies:starburst, galaxies:evolution, galaxies:structure

## 1. Introduction

Searches for the optical counterparts of galaxies detected in the far-infrared have revealed that the most luminous FIR sources are generally associated with peculiar galaxies, most of which are thought to be merging systems (Sanders et al. 1988). However, the ultimate power sources in these "ultraluminous" systems remain as yet unclear; either vigorous star formation or an active nucleus could account for their copious FIR emission. While a number of these systems show emission line diagnostics characteristic of AGN (Sanders et al. 1988), many others show emission line ratios (Leech et al. 1989) and radio source sizes (Condon et al. 1991) more representative of compact central starbursts. The presence of very high density gas in the central regions of some ultraluminous IR galaxies (e.g., Solomon et al. 1992) provides an environment conducive to rapid star formation. Numerical work has shown that mergers are very effective at triggering an inflow of disk gas into the nuclear regions, producing such large concentrations of gas (e.g., Barnes & Hernquist 1991) which could fuel starburst activity (Mihos, Richstone, & Bothun 1992).

However, while the picture of merger-driven starbursts can qualitatively explain many of the properties of the ultraluminous infrared galaxies, a number of details remain unresolved. Perhaps most important is the discrepancy between the derived ages and gas depletion times for the starbursts and the dynamical time of the merging process. Typical ages derived for starburst events are $\sim 10^7 - 10^8$ years (Wright et al. 1988, Bernloehr 1993), much shorter than the merging timescale, $\sim 10^9$ years (e.g., Barnes 1992, Hernquist 1992, 1993a). Furthermore, the most luminous FIR galaxies seem only to be associated with galaxies in the final stages of merging (Sanders et al. 1988, Melnick & Mirabel 1990, Majewski et al. 1993), suggesting that galaxies are stable against such violent starbursts during $> 90\%$ of the interaction history. In order to account for this long period of relative inactivity, either the gas inflow which triggers a starburst must be delayed until the final stage of the merger, or star formation must be inhibited in any massive nuclear gas concentrations which develop during the early stages of a merger.

Models of merger-driven gas inflow and subsequent starbursts have yielded mildly ambiguous results on these issues. Simulations by Barnes & Hernquist (1991) show that disk galaxy collisions are able to drive strong radial inflows of gas shortly after the initial impact, suggesting that induced starbursts should begin well before the merger is complete. Mihos et al. (1992, 1993) used models which included star formation to show that unless galaxies merge quickly, star formation depletes the gas and limits the starburst intensity during the final merger. Therefore, while the models of Mihos et al. can account for many properties of merging galaxies, they were unable to explain the emission from the most extreme objects by appealing solely to a starburst.

In an effort to explore further the triggering of starbursts, we have used numerical simulation to model star formation in a variety of merging systems. In this *Letter* we present the results of a subset of our calculations to examine the sensitivity of the induced starbursts to the structural properties of the progenitor galaxies. In particular, we find that the inner structure of galaxies determines to a great extent the formation and evolution of starbursts in mergers. Galaxies with dense central bulges prove stable against strong starbursts until the final merger, providing a natural explanation for the discrepant merging and starburst timescales. The intense starbursts in these models can account for the extreme IR emission from ultraluminous IR galaxies without the need to invoke an active nucleus. Conversely, bulgeless galaxies are prone to moderate starburst activity at a much earlier phase, and do not correspond well to observed merging galaxies or ultraluminous IR galaxies. An understanding of the details of bulge formation in galaxies is therefore crucial for interpreting the evolution of galaxies.

## 2. Numerical Technique

We model the evolution of the merging galaxies using a hybrid $N$-body/hydrodynamics code (TREE-SPH; Hernquist & Katz 1989), including the effects of star formation (Mihos & Hernquist 1994b). The techniques employed in our calculations have been described in detail elsewhere and will be summarized only briefly here. We use a tree algorithm (Barnes & Hut 1986; Hernquist 1987) to compute the gravitational forces on the collisionless and gaseous components of the galaxies, and smoothed particle hydrodynamics (SPH; e.g., Lucy 1977; Gingold & Monaghan 1977) to model the evolution of the disk gas. The gas is evolved according to hydrodynamic conservation laws, including an artificial viscosity to capture shocks. For expediency, the gas is forced to be isothermal, at a temperature of $10^4$ K. Because of the rapid cooling times of dense disk gas, increases in the thermal energy of the gas from shocks are quickly radiated away (e.g., Hernquist 1989, Barnes & Hernquist 1991), so that the use of an isothermal equation of state scarcely affects the dynamics of the dense gas (Barnes & Hernquist 1994). The evolution of tidal tails in our models will be compromised to some extent, as this low density gas is not allowed to cool adiabatically and is likely, therefore, to be overly diffuse. However, since the starbursts in our models are confined to regions of high gas density in the centers of the galaxies, details of the tidal tails at large radius ($> 50$ kpc) will not affect derived starburst properties.

The galaxies used in our calculations are constructed using the method of Hernquist (1993b), in which particles are distributed in space according to input structural parameters and are given velocities initialized from moments of the Vlasov equation, assuming Gaussian velocity distribution functions. Each galaxy consists of a self-gravitating disk and halo, with an optional central bulge. In the units employed here, disks have an exponential radial scale length $r_s = 1$ and total mass $M_d = 1$. Halos are represented by truncated isothermal spheres, having density profiles $\rho_h(r) \propto \exp(-r^2/r_t^2)/(1 + r^2/\gamma^2)$, with $\gamma = 1$, $r_t = 10$, and total mass $M_h = 5.8$. Optional bulges are described by an oblate generalization of the potential-density pair proposed by Hernquist (1990) for spherical galaxies and bulges, and have a major axis scale length $a = 0.2$, a minor axis scale length $c = 0.1$, and total mass $M_b = 1/3$. The total number of collisionless particles in each component is $N_d = 32768, N_h = 32768,$ and $N_b = 8192$. The disk gas in each galaxy, comprising 10% of the total disk mass, is initially represented by 16384 gas particles distributed in a manner similar to that of the disk stars, but with smaller vertical width. Scaling the model units to match physical values, unit length is 3.5 kpc, unit mass is $5.6 \times 10^{10}$ M$_\odot$, and unit time is $1.3 \times 10^7$ years.

To include the effects of star formation, we note that observations of nearby galaxies suggest that the star formation rate in disks can be reasonably parameterized as a function of the local gas density (i.e. Schmidt 1959): SFR (M$_\odot$ yr$^{-1}$ pc$^{-3}$) $\propto \rho_{gas}^n$, where $n$ has been empirically derived to lie in the range $1 < n < 2$ (e.g., Berkhuijsen 1977; Kennicutt 1989 and references therein). We use a Lagrangian form of this Schmidt law, calculating the star formation rate per unit mass in a gas particle as $\dot{M}/M = C \times \rho_{gas}^{\frac{1}{2}}$, with the constant of proportionality $C$ chosen such that an isolated disk galaxy forms stars at a global rate of 1 M$_\odot$ yr$^{-1}$. Integrated over volume, this prescription gives a good representation of a Schmidt law with index $n \sim 1.5$. Once star formation rates are calculated for each particle, a fraction of the energy feedback is given to the random motions of surrounding gas particles, thereby mimicking the effects of energy input to the ISM from supernovae and stellar winds from massive stars.

The depletion of gas and formation of new stars is handled by modeling the gas using hybrid gas/young star particles, which are gradually converted from gaseous to collisionless form. Each particle is characterized by both a total mass and a gas mass; the gravitational force on a particle is calculated using its total mass, while the gas mass is used to calculate the hydrodynamic forces and properties of the gas.

Through star formation, the gas mass is slowly reduced while the total mass remains constant, thereby mimicking the gas depleting effects of star formation. When the gas mass of a hybrid particle drops below 5%, it is converted to a pure collisionless particle, and thus becomes a good tracer of the newly formed population of stars.

The galaxies are initially separated by 30 length units and placed on parabolic orbits whose pericentric distance at first passage would be 2.5 length units for the ideal Keplerian orbit. While much closer encounters may yield significantly different results, such nearly head-on collisions must be very rare, and not typical of galaxy mergers in general. The geometry of the encounter is described by the inclination of the disks to the orbital plane, $i_1$ and $i_2$, and their arguments of periapse, $\omega_1$ and $\omega_2$ (e.g., Toomre & Toomre 1972). We have run a sequence of calculations exploring the parameter space of inclinations. For brevity, the models shown here describe a merger involving one perfectly prograde galaxy ($i_1 = 0$, $\omega_1$ undefined), and one highly inclined galaxy ($i_2 = 71$, $\omega_2 = 30$) (cf. Barnes & Hernquist 1991; Hernquist 1992, 1993a model 3). However, the major conclusions described here apply to all the models we have run to date.

## 3. Results

Figure 1 (Plate 1) shows the morphology of the star forming gas during the merger of two bulgeless disk/halo galaxies, while Figure 2 shows the evolution of the global star formation rate in the system. Prior to their first close passage, the disks remain relatively unperturbed and show no increase in star forming activity until near perigalacticon ($T \sim 20$). The strong tidal forces at this time greatly deform the disks, leading to the formation of tidal tails and bridges, as well as a prominent bar in the central portion of each galaxy. The gaseous bars tend to lead the stellar bars in each disk by a few degrees (Barnes & Hernquist 1991); the subsequent torque on the gas from the disk stars drives a rapid inflow of gas into the central regions of the galaxies. The resulting high gas densities in these regions trigger strong starburst activity in the disks as they reach their widest separation before finally merging. The star formation rate in these starbursts peaks at 20–30 times the initial value, with a factor of 10 increase in the star formation rate sustained for $\sim 150$ Myr. This rapid conversion of gas to stars depletes the starburst of its fuel, causing the starburst to die out *before* the final merger at $T \sim 70$.

---

**1 MB POSTSCRIPT GREYSCALE or COLOR GIF
AVAILABLE UPON REQUEST FROM hos@lick.ucsc.edu**

---

**Figure 1** – Star forming morphology of two merging disk/halo galaxies. Dark shading represents most intense star formation; a logarithmic scale has been used to emphasize faint structure. Time is shown above each map, with unit time being $1.3 \times 10^7$ years. Due to saturation in the image intensity at the highest levels of star forming activity, it is difficult to intercompare absolute levels of star formation in this Figure; for example, star formation in the dense knots at T=36 is stronger than the entire disk star formation rate at T=12. Absolute star formation levels can be examined in Figure 2.

When the galaxies do finally merge, the ensuing starbursts are extremely weak, only a factor of a few over their pre-interaction levels. The reason behind this inactivity is the lack of gas available to fuel a starburst. Figure 3 shows the evolution of total gas mass in the system as a function of time. Well over half the gas initially in the galaxies is consumed during the strong, widely separated starburst phase; since some gas is also launched to large radius in the tidal tails, only a small amount is left in the center of the merging galaxies. Subsequent infall from gas in the tidal tails is limited and the merger remnant evolves only passively from this point onward.

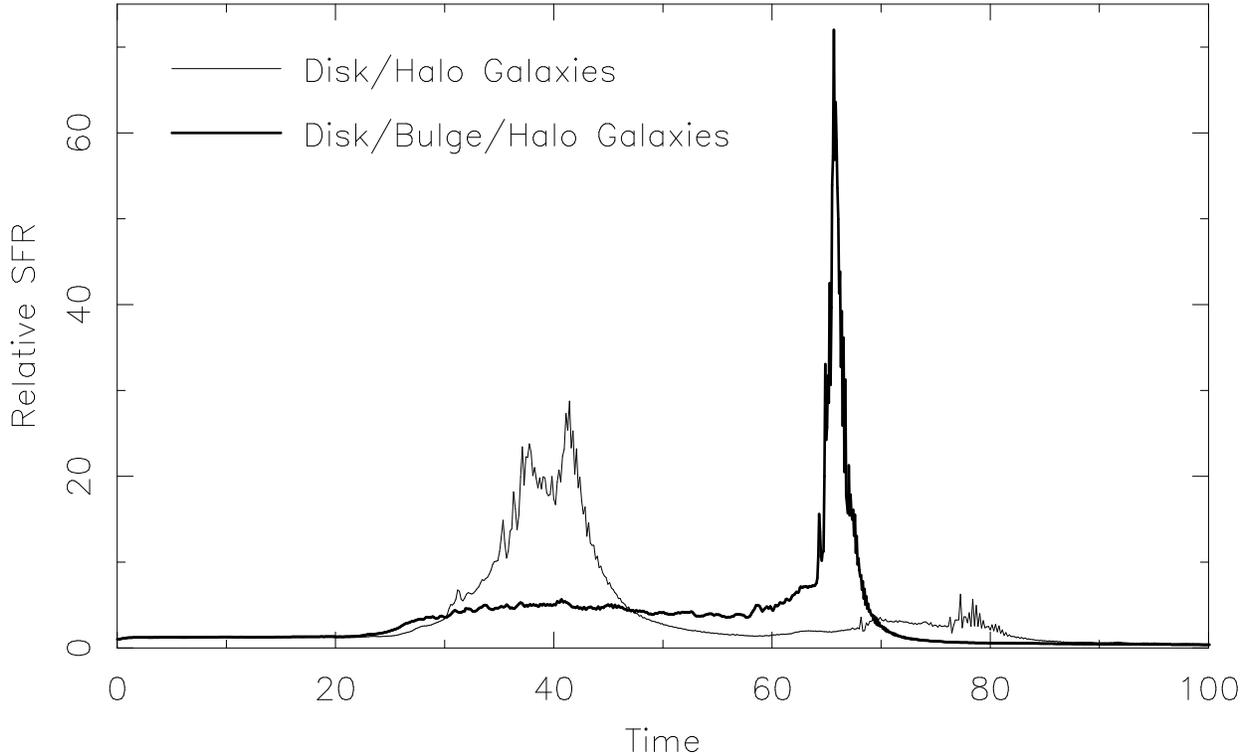

**Figure 2** – Evolution of the global star formation rate in galaxy merger events.

In order to determine the effects of different disk geometries, we ran a series of models with prograde, retrograde, and tilted disks; while the details of the evolution varied in each model, the overall picture remained similar. Regardless of orientation, starbursts develop in the disk galaxies while they are still widely separated, exhausting the gas before the final merger. With little gas remaining to fuel star formation, the ensuing starbursts in the merger remnants are very weak. The lack of dense gaseous cores and strong star formation in the merger remnants makes these models a poor match to the observed properties of ultraluminous IR galaxies. Accordingly, these rare galaxies *do not* seem to result merely from galaxy mergers of special orbital geometry; other factors must act to trigger the onset of such strong activity.

Turning now to galaxy models which include a central bulge component, Figure 4 (Plate 2) shows the star forming morphology during a merger of two disk/bulge/halo galaxies, with identical geometry and orbital parameters as the merger shown in Figure 1. As before, the disks reach their first close approach near $T \sim 20$, and then separate before finally falling back upon one another and merging around $T \sim 70$. However, unlike the bulgeless disks, these disks do not develop a pronounced bar after the initial encounter; instead, they sport two-armed spiral features (Hernquist 1993a) which are less effective at driving inflows of gas to the galaxy centers. As a result, the starbursts ensuing after first close approach are much less intense than those in bulgeless disks, exhibiting increases in the global star formation rate of factors of only a few (Figure 2). These weak starbursts do not severely deplete the disk gas (Figure 3), setting the stage for greater activity during the final merger.

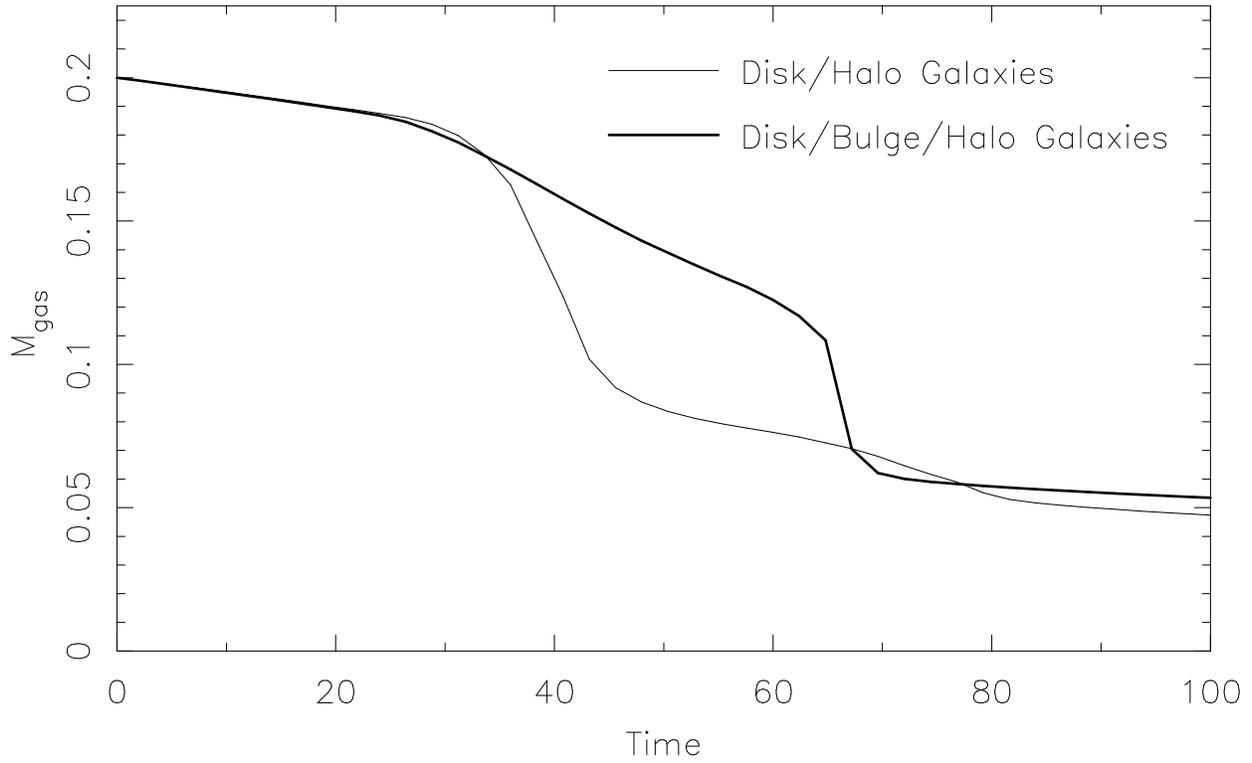

**Figure 3** – Evolution of the total gas mass in galaxy merger events.

When the galaxies finally coalesce, they have fully twice the gas content of the bulgeless disks at a similar time, and the strong dissipation associated with the merger drives this gas into a compact structure at the center of the remnant. The extremely high gas density of this central blob triggers a powerful starburst, whose maximum star formation rate peaks at over seventy times the pre-interaction rate. Due to rapid depletion of the star forming gas, this starburst is very short-lived, lasting for only $\sim 50$ Myr. After the burst is completed, the galaxy evolves quiescently as the remaining gas has been dispersed in a diffuse manner around the remnant and in the tidal tails. The old disk and bulge populations in these types of mergers show $r^{\frac{1}{4}}$ law mass profiles over a large range of radius, typical of present-day ellipticals (Hernquist 1993a). With the young centrally concentrated population and $\sim 10^9$ $M_\odot$ of remaining neutral gas, these remnants may well be classified as E peculiar galaxies after several Gyr. Our models, therefore, provide a clear link between the ultraluminous infrared galaxies and the formation of at least some fraction of the local elliptical galaxy population.

---

**1 MB POSTSCRIPT GREYSCALE or COLOR GIF
AVAILABLE UPON REQUEST FROM hos@lick.ucsc.edu**

---

**Figure 4** – Star forming morphology of two merging disk/bulge/halo galaxies. Dark shading represents most intense star formation; a logarithmic scale has been used to emphasize faint structure. Time is shown above each map, with unit time being $1.3 \times 10^7$ years.

Again, we have run a number of encounters with differing disk geometries to assess the star forming properties of merging disk/bulge/halo galaxies. We find similar results in all these models – a period of mild starburst activity after the first passage, followed by short-lived, extremely powerful starbursts during the final merger. In fact, the star formation rate increased by over a factor of 100 in one encounter involving a prograde/retrograde disk pair. In general, therefore, it appears that central bulges can partially stabilize disk galaxies against gas inflow and inhibit strong starbursts during the majority of a merger. This result echos previous simulations of disk galaxies accreting dwarf companions (Mihos & Hernquist 1994a), in which it was found that a central bulge stabilized the disk against very strong starbursts. However, in the case of major mergers the subsequent violent merging of the galaxies overwhelms the kinematic support provided by the bulge, resulting in an extreme starburst in the center of the remnant.

## 4. Discussion

While limited in their coverage of parameter space, our models shed new light on the nature of starburst activity in merging galaxies. In particular, they provide a solution to the problems of gas depletion and disparate merging/starburst timescales. Dense central bulges stabilize the disk against bar formation after the initial passage, suppressing gas inflow and starbursts until the final stages of a merger. Galaxy models which lack a dense bulge experience a much earlier starburst phase, and display properties unlike those observed in nearby interacting and merging galaxies. The structure of the merging galaxies, more so than the orbital geometry, determines the nature of the starburst event. The calculations presented here may be compared to those of Barnes & Hernquist (1994), who employed bulges much more diffuse than those in our models. The diffuse bulges provide limited kinematic support during the early phase of the merger, leading to an evolution of the gas component much like that in our bulgeless models. Therefore, it appears that bulges must be sufficiently dense in order to stabilize disk galaxies against an early starburst phase.

Merging galaxies which contain central bulges also display many of the prominent features of the ultraluminous IR systems. At the peak of their starburst phase, these model galaxies are characterized by a very small separation and high gas content. The hundred-fold increase in star formation rate can account for the levels of IR emission from the most extreme objects without invoking an active nucleus, and the timescale of the starburst is consistent with that derived from gas depletion arguments or spectral synthesis techniques. Furthermore, the sudden onset and rapid decline of the starbursts indicates that the dominance of close mergers in galaxy samples selected on the basis of extremely high infrared emission (e.g., Sanders et al. 1988, Majewski et al. 1993) is simply a selection effect resulting from the fact that only during this brief phase in a merger are such high levels of activity achieved.

Previous studies of ultraluminous IR galaxies have been unable to determine whether the low space density of these objects is a result of special and rare initial conditions (i.e. very high gas content or special orbital geometry), or if they represent a very brief but common phase of galaxy mergers. The results presented here favor the latter interpretation. Regardless of disk inclination, all the model galaxies with bulges developed a very brief, intense starburst. The properties of the disk galaxies, as characterized by the rotation curve and gas content, are typical of present-day disk galaxies, and do not represent "special" initial conditions. In general, therefore, mergers of disk galaxies with dense bulges should produce extremely high levels of star formation for a brief period, appearing as ultraluminous IR sources. Since galaxy mergers were probably more common at early epochs, these ultraluminous galaxies may comprise a large fraction of the galaxy population at high redshift, and, furthermore, may represent a brief but important phase in the formation of many low redshift elliptical galaxies.

Our results indicate that the structure of the progenitor galaxies determines to a great extent the formation and evolution of starbursts in merging galaxies. If dense bulges formed before or concurrently with galactic disks, then merging disk galaxies at high redshift should be relatively stable against gas inflow and not experience strong starbursts until the final stages of merging, at which point ultraluminous starburst activity is triggered. If, on the other hand, bulges were incorporated into galaxies after the formation of the disks, galaxies merging at high redshift should show properties more indicative of the bulgeless galaxy models shown here, perhaps displaying high levels of activity while still widely separated, and a relatively quiescent merging period. Determining the era of bulge formation in disk galaxies will therefore be of crucial importance in understanding the role starburst activity plays in the evolution of galaxies.

This work was supported in part by the San Diego Supercomputing Center, the Pittsburgh Supercomputing Center, the Alfred P. Sloan Foundation, NASA Theory Grant NAGW–2422, the NSF under Grants AST 90–18526 and ASC 93-18185, and the Presidential Faculty Fellows Program.